\documentstyle[aaspp]{article}

\begin{document}

\def\LSUN{\rm L_{\odot}}
\def\MSUN{\rm M_{\odot}}
\def\RSUN{\rm R_{\odot}} 
\def\MSUNYR{\rm M_{\odot}\,yr^{-1}}
\def\MDOT{\dot{M}}

\newbox\grsign \setbox\grsign=\hbox{$>$} \newdimen\grdimen \grdimen=\ht\grsign
\newbox\simlessbox \newbox\simgreatbox
\setbox\simgreatbox=\hbox{\raise.5ex\hbox{$>$}\llap
     {\lower.5ex\hbox{$\sim$}}}\ht1=\grdimen\dp1=0pt
\setbox\simlessbox=\hbox{\raise.5ex\hbox{$<$}\llap
     {\lower.5ex\hbox{$\sim$}}}\ht2=\grdimen\dp2=0pt
\def\simgreat{\mathrel{\copy\simgreatbox}}
\def\simless{\mathrel{\copy\simlessbox}}

\title{Resonance line-profile calculations based on
hydrodynamical models of cataclysmic variable winds.}

\vspace{1.cm}
\author{ Daniel Proga} 
\vspace{.5cm}
\affil{LHEA, GSFC, NASA, Code 662, Greenbelt, MD 20771} 
\affil{Presently at JILA, University of Colorado, Boulder, CO 80309-0440;
proga@colorado.edu}

\author{Timothy R. Kallman }
\vspace{.5cm}
\affil{LHEA, GSFC, NASA, Code 662, Greenbelt, MD 20771; 
tim@xstar.gsfc.nasa.gov}

\author{ Janet E. Drew \& Louise E. Hartley}
\vspace{.5cm}
\affil{Imperial College of Science, Technology and Medicine,
Blackett Laboratory, Prince Consort Road, London, SW7 2BZ, UK;
j.drew@ic.ac.uk, le.hartley@ic.ac.uk}

\begin{abstract}

We describe a method of calculating synthetic line profiles
using a generalized version of the Sobolev approximation.
We apply this method to calculate line profiles predicted
by the models of two-dimensional line-driven winds from luminous disks 
due to Proga, Stone \& Drew. We describe the main properties of the model 
line profiles and compare them with recent HST observations of 
the cataclysmic variable IX Vel.  The model wind consists of a
dense, slow outflow that is bounded on the polar side by a
high-velocity  stream. We find that these two wind components
produce distinct spectral features. The fast stream produces
profiles which show features consistent with observations.
These include the appearance of the classical P-Cygni shape for a range of 
inclinations, the location of the maximum depth of the absorption 
component at velocities less than the terminal velocity, and 
the transition from net absorption to net emission with increasing inclination.
However the model profiles have too little absorption or emission
equivalent width compared to  observed profiles. 
This quantitative difference between our models and observations is not
a surprise because the line-driven wind models predict a mass loss
rate, mostly due to the fast stream, that is lower than the rate required  
by the observations. We note that the model profiles exhibit
a double-humped structure near the line center which is not 
echoed in observations.
We identify this structure with a non-negligible redshifted 
absorption which is formed in the slow component of the wind where 
the rotational velocity dominates over expansion velocity.
We conclude that the next generation of disk wind models, developed for
application to CVs, needs to yield stronger wind driving out to larger disk 
radii than do the present models.

\end{abstract}


\section{Introduction}

Outflows appear to be a common feature of many types of accreting compact
objects. In the case of cataclysmic variables (CVs), key evidence
for outflows comes from P-Cygni profiles of strong UV lines such
as C~IV$\lambda$1549.  But until recently there was no ab initio dynamical 
model of these outflows, and interpretation of data was limited to fitting
observed profiles to synthetic profiles calculated from kinematic
models. This situation has changed with the advent of numerical models
for the hydrodynamics of disk winds, and it is now possible to calculate
synthetic profiles from them. This is what we set out to do in this
paper.

The International Ultraviolet Explorer (IUE) satellite (Boggess et al. 1978)  
first discovered that CVs exhibit P~Cygni type line profiles 
(Heap et al. 1978). These were mainly identified in low resolution 
($\lambda/\Delta \lambda \sim 300$) IUE spectra where absorption troughs 
extending bluewards over $\sim 5000$~km~s$^{-1}$ were noted (e.g., Cordova \& 
Mason 1984; Drew 1990).   This phenomenon is attributed to high-velocity 
winds by analogy with OB stars (e.g., Krautter et al. 1981).
The affected line transitions are without exception, C~IV~$\lambda$1549 and 
N~V~$\lambda$1240 and, in many instances, Si~IV~$\lambda$1397. Because the 
maximum observed wind speeds are often comparable with the escape speed from 
a WD photosphere, it has been presumed that CV winds are disk winds 
originating from close to, if not on, the accreting WD (Cordova \& Mason 
1982).

The particular varieties of CV that most clearly exhibit wind features
are the non-magnetic systems in the high state: the dwarf novae (DN) in 
outburst and nova-like variables (NLs). The main difference between these CV 
sub-types is in their secular light curves: DN undergo brightness 
increases (outbursts) of several magnitudes at quasi-regular intervals, 
whereas NLs are more stable objects staying always in a high state. The 
outbursts in DN are understood as rapid changes in the physical structure of 
the accretion disk -- the instabilities being related to changes in the mass 
accretion rate (Cannizzo 1993).
 
The observed low-resolution UV profile shape depends sensitively 
on the inclination angle of the system. Observed profiles range from
broad shortwardshifted absorption, with or without redshifted emission,
in cases of low-to-intermediate inclination angles ($\leq 60^o$), to 
Doppler-broadened emission profiles in luminous CVs viewed almost edge on.
This behavior is consistent with line formation mainly by resonant
scattering in a somewhat bipolar disk wind (Drew \& Verbunt 1985, 
Drew 1987).

The UV observations have also shown that CV mass loss signatures  
can be very sensitive to luminosity. During declines of DN
from maximum light, the UV continuum decays by a factor of $\simgreat 2$;
the blueshifted absorption prominent at the maximum light often
nearly vanishes during the decline (e.g., Woods et al. 1992). Analogous 
behavior has been noted in some NLs as their brightness level varies (e.g., 
Mason et al. 1995). 

Other properties of the UV resonance line profiles that are likely related 
to winds  have been observed in the small fraction of highly-inclined
non-magnetic CV that exhibit deep continuum eclipses.  For example, weak UV 
line eclipses or even line brightenings were found at times of
continuum eclipse in UX~UMa (King et al. 1983), RW~Tri (Cordova \& 
Mason 1985) and OY~Car (Naylor et~al. 1988).  These weak 
line eclipses indicate either a size of emission line region larger than the 
secondary star, or an out-of-eclipse line profile that is a spatially 
unresolved superposition of a net absorption component on a stronger emission 
feature (Drew 1987).  A further observed property of the emission lines in 
eclipsing systems is the narrowing of the UV resonance lines during the 
eclipse (OY~Car: Naylor et al. 1988, V347 Pup: Mauche et al. 1994, and 
UX~UMa: Mason et al. 1995). 

Research on the supersonic outflows in CVs has nearly closed a circle 
in science: from observations and simple phenomenological models through 
sophisticated kinematic models to first generation of dynamical models. There 
have been many successes in this process; for example, the kinematic models 
developed by Drew (1987), Mauche \& Raymond (1987) and later by Shlosman, 
Vitello \& Mauche (1996) and by Knigge \& Drew (1997) provide a point of 
comparison for dynamical theories of outflows.  Dynamical models appealing
to radiation pressure driving have met with qualitative success in explaining 
some observed disk wind properties (Pereyra, Kallman \& Blondin 1997; 
Proga, Stone \& Drew 1998, 1999, hereafter PSD~98 and PSD~99; Proga 1999; 
Feldmeier \& Shlosman 1999; Feldmeier, Shlosman \& Vitello 1999; 
Pereyra, Kallman \& Blondin 2000).  Important outcomes of the hydrodynamical 
simulations were (i) the mass loss rate's practical independence of the 
radiation field geometry -- it is first and foremost a function of total 
system luminosity, (ii) the discovery of unsteady flow in models where the 
disk radiation field dominates over any contribution from the central 
accreting object.  To bring closure of the scientific circle even
nearer, the main objective here is to calculate synthetic line profiles 
directly from the line-driven disk wind models of PSD~99 and to critically
compare them with relevant high quality UV observations.

Our focus shall be on assessing the gross properties 
of the line profiles - such as the strength and shape of the blueshifted
absorption.  In Section~2,
we briefly review the past work in the field 
of winds in CVs and then describe our method of calculating line profiles, 
using a generalized Sobolev approximation.  
The results are presented in Section~3. 
In Section~4, we discuss key aspects of the synthetic line profiles and make
comparisons with recent observations.

\section{Method}

Recently numerical hydrodynamic models of 2.5-D, time-dependent radiation 
driven disk winds have been constructed for application to, in the first
instance, CV disk winds (Pereyra, Kallman \& Blondin 1997; PSD~98;
PSD~99; Proga 1999).   The primary outcome of these studies is the 
confirmation that radiation-pressure due to spectral lines can produce a 
supersonic, biconical outflow from an accretion disk in nMCVs.  
Additionally, PSD~98 and PSD~99 found that regardless of the radiation 
geometry,  the two-dimensional structure of the wind consists of a
dense, slow outflow that is bounded on the polar side by a
high-velocity  stream (respectively `slow wind' and `fast stream', 
for short; see e.g., Figure 2 in PSD~99).

Applying PSD~98 and PSD~99's dynamical models to CVs one 
finds that radiation driving can produce disk winds consistent with the 
following observed properties of CV winds:\\
(1) the flow is biconical rather than equatorial as required by the
absence of blueshifted line absorption from the spectra of eclipsing
high-state nMCV,\\ 
(2) the wind terminal velocity is comparable to the escape velocity from the 
surface of the WD,\\  
(3) the spectral signatures of mass loss show a sharp cut-off as the total 
luminosity in DN declines away from maximum light through the regime 
theoretically identified as likely to be critical.\\ 
Additionally,  line-driven disk wind models may explain the highly unsteady 
and continuously variable nature of the supersonic outflows in the NL binaries
BZ~Cam and V603 Aql (Prinja et al. 2000a; b).  The presence of a slow, dense 
transition 
region between disk photosphere and outflow in V347~Pup and UX~UMa 
(Shlosman, Vitello \& Mauche 1996; Knigge \& Drew 1997) 
may also be accommodated within these same models.
These successes of the line-driven wind model are encouraging and provide the 
motivation for its closer examination.

\subsection{Line-profile Calculations}

We restrict our attention to the case of a representative UV  
resonance transition of a light ion such as C~IV or Si~IV.  The 
assumed abundance and atomic data are appropriate to the C~IV 1549\AA\ 
transition treated as a singlet.  The profiles synthesized on this basis 
allow us to address the main issues of this paper.

As in the hydrodynamical calculations, we apply  a generalized Sobolev 
approximation (Sobolev 1957; Rybicki \& Hummer 1978, 1983) 
in treating spectral line transport through
a wind. Our calculations are similar to those by Shlosman \& Vitello (1993).
The main differences are (i) we base our calculation  on 
numerical hydrodynamical wind models whereas Shlosman \& Vitello 
used kinematic wind models
(ii) we allow both the white dwarf and the accretion disk
to radiate but neglect the boundary layer 
whereas Shlosman \& Vitello considered
the disk, white dwarf and boundary layer as a source of radiation
(iii) we assume the wind ionization state whereas Shlosman \& Vitello
calculated the photoionization of the wind in an optically thin 
approximation.
A minor technical difference between our approach and that of Shlosman \& 
Vitello is that we use spherical polar coordinates ($r, \theta, \phi$)
while they used cylindrical coordinates ($R, Z, \phi$).  
We form  the three-dimensional computational grid
by rotating the spatial $(r,\theta)$ grid in the $\phi$ direction in the
same way as in the hydrodynamical calculations used as input..

Below is a summary of our method.
The radiation propagates in an axisymmetric system with a fully
three-dimensional velocity field $(v_r, v_\theta, v_\phi)$.
We define the monochromatic specific luminosity as 
\begin{equation}
L_\nu \equiv \int I_\nu(\hat{n})dA, 
\end{equation}
where $I_{\nu}(\hat{n})$
is the intensity at an arbitrary point along the unit vector $\hat{n}$ 
pointing toward the observer. 
We define a surface element, $dA$ in the plane of the sky
normal to $\hat{n}$.
In the Sobolev approximation,
it is possible to analyze the radiation transport
as if all the contributions to the formation of a spectral line
at any given frequency occurred locally. In particular,
along a given ray the intensity at frequency $\nu$ does not change
at all, except at certain discrete resonance points, where the 
material has just the right Doppler shift to allow it to absorb or
emit. These points occur wherever the line-of-sight
velocity $v_l\equiv {\bf v \cdot} \hat{n}$ satisfies the resonance
condition:
\begin{equation}
(\nu-\nu_0)/\nu_0=v_l/c.
\end{equation}
To calculate the observed line profile, we need to compute the intensity
that emerges from the resonance region,
\begin{equation}
I^{emg}_\nu=I^{inc}_\nu \exp\left(-\sum_{i=1}^N \tau_i\right)+ \sum_{j=1}^N
s_j[1-\exp(-\tau_j)]\exp\left(-\sum_{i=1}^{j-1}\tau_i\right), 
\end{equation}
where $s$ is the line source function and 
$I^{inc}$ is the incident intensity due to a disk or stellar surface, 
The incident intensity is non-zero at a point $\bf r$ if $\hat{n}$ 
originates on the disk or star surface. 
The factor
$\exp(-\sum_{i} \tau_i)$  accounts for non-local
effects  in a generalized Sobolev approximation:
the term $\tau_i$ is the line optical depth at a possible intervening
resonance velocity surface between the point $\bf r$ and the observer.  
In the original version of the Sobolev approximation, the flow
velocity along a line of sight is assumed to be monotonic and 
thus only one resonance
surface is considered, i.e., $N=1$ or 0 and $\exp(-\sum^{j-1}_{i=1} \tau_i)=1$.
Following Rybicki \& Hummer (1983), we will refer to the first term
on the RHS of eq. 3 as the surface contribution and the second
term as the volume contribution to the emergent intensity.
These two contributions can be treated separately.

Rybicki \& Hummer (1983) showed that the area integral of the terms
in the emergent intensity which are linear in the values of the source 
functions can be replaced by a volume integral,
\begin{equation}
L^v_\nu=\int j_\nu({\bf r}) p_\nu({\bf r},\hat{n}) {\bf dr}^3,
\end{equation}
where $j_\nu$ is the emission coefficient, 
and $p_\nu$ is the directional escape probability for 
a photon emitted at point $\bf r$ in direction $\hat{n}$ and frequency $\nu$.
Equation~4 represents the scattered emission of the resonance line.
In the Sobolev approximation, the emission coefficient can be 
expressed as
\begin{equation}
j_\nu= k({\bf r}) s({\bf r})\delta(\nu -\nu_0 -\nu_c v_l/c),
\end{equation}
where $k$ is the line opacity and 
$\delta$ is the Dirac delta function. The escape probability
can be written as
\begin{equation}
p_\nu=\frac{1-\exp[-\tau({\bf r},\hat{n})]}{\tau({\bf r},\hat{n})}
\exp\left(-\sum^{j-1}_{i=1} \tau_i\right).
\end{equation}
The optical depth can be approximated
by the Sobolev optical depth as
\begin{equation}
\tau({\bf r},\hat{n})=\frac{kc}{\left|d v_l/dl\right|},
\end{equation}
where $d v_l/dl$ is the velocity gradient along $\hat{n}$ and
may be written as
\begin{equation}
\frac{dv_l}{dl}=Q~\equiv~ \sum_{i,j}\frac{1}{2}\left(\frac{\delta v_i}{\delta r_j}
+\frac{\delta v_j}{\delta r_i}\right)n_in_j=\sum_{i,j}e_{ij}n_in_j,
\end{equation}
where $e_{ij}$ is the symmetric rate-of-strain tensor.
Expressions for the components of $e_{ij}$ in the spherical polar coordinate
system are given in Batchelor (1967).  To compute the line optical depth,
we use the atomic data for the C~IV~$\lambda$1549 transition (i.e.
we set the oscillator strength to $0.1911$).

To calculate the line opacity, we need to calculate 
the number density of the absorbing and scattering ions:
\begin{equation}
n_{ion}({\bf r})=\frac{\rho({\bf r})}{\mu_p m_p} A~\xi_{ion},
\end{equation}
where $\rho$ is the mass density, $m_p$ is the proton mass,
$\mu_p$ is the mean molecular weight per proton, 
$\xi_{ion}$ is the ion fraction, and $A$ is the abundance of 
a given element with respect to hydrogen assumed to be $3\times 10^{-4}$.
We also assume that the ionization stage of our typical element is
constant and set both $\mu_p$ and $\xi_{ion}$ to unity.
Thus the number density of the absorbing and scattering ions depends 
on position only through the mass density.

Although we take into account non-local coupling caused by non-radial velocity
fields while calculating the escape probability we ignore this
coupling while calculating the source function. In other words, 
we calculate the source function assuming that only a single resonance 
velocity surface exists -- this is known as the disconnected approximatio n 
(Marti \& Noerdlinger 1977).  We express the source function as
\begin{equation}
s({\bf r})=\frac{<\beta_{\nu_0} I_{\nu_0}>_{\Omega}}{<\beta_{\nu_0}>},
\end{equation}
where $<\beta_{\nu_0} I_{\nu_0}>_\Omega$ is 
the angular averaged intensity from the radiative source (a disk,
a star or both) weight by the escape probability defined as 
\begin{equation}
<\beta_{\nu_0} I_{\nu_0}>_\Omega=\frac{1}{4\pi}
\oint_{\Omega} p_{\nu_0} I^{inc}_{\nu_0} d\Omega,
\end{equation}
where $\Omega$ is the solid angle subtended by the disk and star.
The term $<\beta_{\nu_0}>$, in eq. 10, 
is the angular averaged escape probability 
defined as
\begin{equation}
<\beta_{\nu_0} >=\frac{1}{4\pi}\oint_{4\pi} p_{\nu_0} d\Omega.
\end{equation}
We use the disconnected approximation  because
it is simple to implement in the code and the resultant equations
can be solved relatively quickly. We note that the complications resulting 
from implementing fully the non-local effects within the Sobolev approximation
spoil the simplicity of the Sobolev approximation so it is better
to use another method to treat the line transport, such as a Monte
Carlo method that allows an exact solution.
At this stage, where we calculate 
line profiles based on  the hydrodynamical models for the first
time,  an approximate simple method suffices to illustrate the 
gross properties of the profiles. 
We also note that we  do not use the thermal part of the source function 
because it is negligible at the temperature we consider.

To calculate the source function at any location, it is first
necessary to specify the disk and star emergent radiation 
intensities.  To do this we adapt the formalism of PSD~99 (see also 
PSD~98). For the frequency-integrated disk intensity of the disk at
radius $r_D$, we use the $\alpha$-disk form (e.g., Pringle 1981):
\begin{eqnarray}
\rule{0in}{3.0ex}
I_{D}(r_D) & =~\frac{3 G M_\ast \MDOT_a}{8 \pi^2 r_\ast^3} 
\left\{ \rule{0in}{3.0ex}  \frac{r_\ast^3}{r_D^3}\left(1 - 
\left( \frac{r_\ast}{r_D}\right)^{1/2}\right) \nonumber \right.~~~~~~~~~~~~~\\
 & \left. +\frac{x}{3\pi}\left(\arcsin \frac{r_\ast}{r_D} - 
\frac{r_\ast}{r_D} \left(1 - 
\left(\frac{r_\ast}{r_D}\right)^2\right)^{1/2}\right) \right\},
\end{eqnarray}
where $r_D$ is the disk radius, 
$M_\ast$  and $r_\ast$ are the mass and radius of the central star, 
$\MDOT_a$ is the accretion rate through the disk (Shakura \& Sunyaev 1973).
We include the effects of the irradiation of a disk by  a star for
$x>0$ where $x$ is defined as the ratio between the stellar luminosity 
$L_\ast$ and the disk luminosity, $L_D$ (PSD~98).
For radiation from a spherical, isothermal star, the frequency-integrated 
intensity may be written:
\begin{equation} 
I_\ast~=~\frac {L_\ast}{4\pi^2 r_\ast^2}~=~x\frac{G M_\ast \MDOT_a }{8 \pi^2 r^3_\ast}.
\end{equation}
We assume that both the disk and star radiate isotropically.
Then we calculate the intensity emerging from the the disk at a frequency 
$\nu$ and radius $r_D$ as the blackbody intensity, $B_\nu$ at the temperature 
$T_D(r_D)=(\pi I_D(r_D)/\sigma)^{1/4}$.  For the radiation from the
star, the frequency-dependent intensity is given as the blackbody intensity 
at the temperature $T_\ast=(\pi I_\ast/\sigma)^{1/4}$.

We evaluate both integrals, $<\beta_{\nu_0} I_{\nu_0}>_{\Omega}$ and 
$<\beta_{\nu_0}>$, using a modified version the numerical method from 
PSD~99 to evaluate the contribution to the line force due to the disk and 
star based on the integration over a solid angle (see Sect.~3
in PSD~99). In particular, we split the integration of 
$\beta_{\nu_0} I_{\nu_0}$
over $\Omega$  into the integration over the stellar  solid angle, 
$\Omega_\ast$, 
so $I^{inc}_\nu=B_\nu(T_\ast)$  
and the disk solid angle, $\Omega_D$
so $I^{inc}_\nu=B_\nu(T_D)$.
We take into account the effects due
to shadowing of the disk by the star, and occultation of the star by
the disk, by properly defining the limits of integration for each.

Having evaluated the values of the integrand in eq. 4 at hydrodynamical
grid points, we numerically sum up the contribution
to the scattered emission over the volume occupied by the hydrodynamical
computational domain. 
The Dirac delta function in the emission coefficient allows the 
volume integral to be reduced to an integral over $r$ and $\theta$.
For a given point ($r$, $\theta$) to be included in the integral, there must
be at least one angle, $\phi_{res}$ for which the $\delta$ function
does not vanish. As discussed by Shlosman \& Vitello, for each value
of the ($r$, $\theta$) [or equivalently ($R, Z$)] coordinates there
are either two, one or zero solutions for the $\phi_{res}$-value
that satisfy the resonance condition. For each resonance point,  we follow
along $\hat{n}$ and reduce 
the contribution of this resonance point to the scattered emission 
by a factor of $\exp[-\tau_i]$
whenever the ray defined by this resonance point
and $\hat{n}$ crosses another resonance velocity surface.

We calculate the surface contribution from the disk and star, $L_\nu^s$
to the luminosity,
$L_\nu$ at the frequency $\nu$ as 
\begin{equation}
L_\nu^s=\int dA I^{inc}_\nu \exp\left(-\sum_{i=1}^N \tau_i\right).
\end{equation}
This equation represents the direct continuum radiation
from the source.  For each ray from the disk or star surface we follow
along $\hat{n}$ and
reduce the incident intensity by a factor of $\exp[-\tau({\bf r},\hat{n})]$
whenever the ray crosses a resonance velocity surface.

\subsection{Wind Models}

For completeness we briefly describe the key elements, assumptions
and parameters of the hydrodynamical models we will use
to calculate line profiles. See  PSD~99 and PSD~98
for the full description.

In the hydrodynamical models, PSD~99 assumed that an optically-thick,
Keplerian  disk surrounds  a non-rotating white dwarf. 
The disk is flat and geometrically thin in the sense that its
radiation is emitted from the disk mid-plane.  
The disk intensity is assumed to be direction independent.
The main input parameters for the 2.5 dimensional line-driven wind model
are: the mass accretion rate through the disk, and 
the stellar mass, radius and luminosity.
PSD~99 calculated several wind models for various $\MDOT_a$ and $L_\ast$
for $M_\ast=0.6~\MSUN$ and $r_\ast=8.7\times10^8$~cm.
For these stellar parameters, the Keplerian velocity at $r=r_\ast$, 
is $v_o=3017~{\rm km~s^{-1}}$.

To calculate the structure of a wind from the disk, PSD~99
took into account stellar gravity, gas pressure effects, rotational
and radiation forces. PSD~99 held the sound speed constant at 
14~${\rm km~s^{-1}}$ in the isothermal equation of state. PSD~99
used the Castor, Abbott, Klein (1975, hereafter CAK) force multiplier 
to calculate the line-driving force.
In this approximation, a general form for this force  
at a point defined by the position vector $\bf r$ is
\begin{equation}
{\bf F}^{rad,l}~({\bf{r}})=~\oint_{\Omega} M(t) 
\left(\hat{n} \frac{\sigma_e I({\bf r},\hat{n}) d\Omega}{c} \right)
\end{equation}
where $I$ is the frequency-integrated continuum intensity in the direction
defined by the unit vector $\hat{n}$, and $\Omega$ is the solid angle
subtended by the disk and star at the point W. 
The term in brackets is the electron-scattering radiation force,
$\sigma_e$ is  the mass-scattering coefficient for free electrons,
and $M(t)$ is the force multiplier -- the numerical factor which
parameterizes by how much spectral lines increase the scattering
coefficient. In the Sobolev approximation, $M(t)$ is a function
of the optical depth parameter
\begin{equation}
t~=~\frac{\sigma_e \rho v_{th}}
{ \left| dv_l/dl \right|},
\end{equation}
where $v_{th}$ is the thermal velocity. 

PSD~99 adopted the CAK  analytical expression
for the force multiplier as modified by Owocki, Castor \& Rybicki 
(1988, see also PSD~98)
\begin{equation}
M(t)~=~k t^{-\alpha}~ 
\left[ \frac{(1+\tau_{max})^{(1-\alpha)}-1} {\tau_{max}^{(1-\alpha)}} \right]
\end{equation}
where k is proportional to the total number of lines,
$\alpha$ is the ratio of optically thick to optically-thin lines,
$\tau_{max}=t\eta_{max}$ and $\eta_{max}$ is a parameter 
related to the opacity of the most optically thick lines.
The term in the square brackets is the Owocki, Castor \& Rybicki correction
for the saturation of $M(t)$ as the wind becomes optically thin
even in the strongest lines, i.e., 
\begin{displaymath}
\lim_{\tau_{max} \rightarrow 0} M(t)~=~M_{max}~=~
k(1-\alpha)\eta_{max}^\alpha.
\end{displaymath}
PSD~99 adopted $k=0.2$ and imposed an upper limit $M_{max}=4400$
for $\alpha=0.6$.

PSD~99's calculations were performed in spherical polar coordinates $(r,
\theta, \phi)$ with $r=0$ at the center of the accreting star.  
Axial symmetry about the rotation axis of the accretion disk ($\theta=0^o$)
was also assumed.  Thus all quantities were taken to be invariant in 
$\phi$.  The standard computational domain was defined to occupy the radial 
range $r_\ast \leq r \leq 10r_\ast$, and angular range $0 \leq \theta \leq 
90^o$.  The $r-\theta$ domain was discretized into zones -   
the standard numerical resolution consisted of 100
zones in each of the $r$ and $\theta$ directions, with fixed zone size
ratios, $dr_{k+1}/dr_{k}=d\theta_{l}/d\theta_{l+1} =1.05$.

\section{Results of Profile Calculations}

We calculate line profiles for four hydrodynamical models
from PSD~99, model A ($\MDOT_a=10^{-8}~\MSUNYR, x=0$),
             model B ($\MDOT_a=\pi\times10^{-8}~\MSUNYR, x=0$)
             model C ($\MDOT_a=\pi\times10^{-8}~\MSUNYR, x=1$)
             model D ($\MDOT_a=\pi\times10^{-8}~\MSUNYR, x=3$).
This range of models illustrates the dependence of the disk
wind on the disk and stellar luminosities. We note that this sequence
of models is a rising sequence in the total system luminosity (see Table~1).
To examine the effect of viewing angle on the line profiles,
for each disk wind model we compute line profiles for five
inclination angles: $i=~5^o, 30^o, 55^o, 70^o$, and $85^o$. Examining
effects of inclination angle is important because of the 
strong inclination angle dependence observed in CV spectra.
As noted earlier, this inclination dependence has been viewed as the key 
evidence for a biconical wind associated with a disk rather than a spherical 
wind associated with a white dwarf.  Figure~1 shows the model line profiles
obtained after scaling to unit continuum level.  We calculate the line 
profiles at a velocity resolution of $\sim 16.7~{\rm km~s^{-1}}$ that 
corresponds to a spectral resolution of $\sim~0.17~\AA$. 

As found by PSD~98 (see also PSD~99), there are two kinds of flow that might 
arise from luminous accretion disks: a complex weak wind and a strong steady 
wind.  We first describe line profiles for a representative example of each
of these two types of winds in some detail first (specifically model A
in section 4.1 and model C in 4.2).  Then we review the effects of varying 
the disk luminosity and the relative luminosity of the white dwarf (see 
section 4.3).

Table~1 summarizes the main input parameters of the hydrodynamical disk
wind models including the mass accretion rate, $\MDOT_a$, total stellar
luminosity relative to the disk luminosity, $x$ and resultant total system
luminosity, $L_{tot}=L_D+L_\ast=(1+x) L_D$. Additionally, the table
lists  the fraction of continuum intensity at the line
frequency that is due to the disk,
$f_D$, for various inclination angles, and the gross 
properties of the disk winds including the mass-loss rate, $\MDOT_w$, 
characteristic velocity of the fast stream at 10$r_\ast$, $v_r(10r_\ast)$ 
and flow opening angle, $\omega$.  As in PSD~99, we define the flow
opening  angle as an angle between the disk equator and the upper envelope
of the wind  at 10$r_\ast$.

\subsection{A complex weak wind case}

First, we present our results for 
a complex weak wind for which
$\MDOT_a=10^{-8}~\MSUNYR$ and the white dwarf is assumed to be dark $x=0$
(model~A).
The gross properties of this wind model are:
the wind mass loss rate $\MDOT_w~=~5.5\times10^{-14}~\MSUNYR$, 
the characteristic
velocity at $10~r_\ast$ $v_r(10~r_\ast)= 900~{\rm km~s^{-1}}$ and
flow opening angle, $\omega=50^o$ (see Table 1 here and table~1 in PSD~99).
We note that $v_r(10~r\ast)\sim 1/3~v_o$.  

The five top panels of Figure~1 (i.e. figs. 1a-1e) show the profiles
for the model~A as a function of $i$. The most obvious change
in the profiles caused by viewing the disk at different
angles is the weakening of the absorption and the strengthening
of the scattered emission as inclination angle increases.
This strong inclination angle dependence of the line
profile is to be expected for a bipolar flow.  
The width of the line also increases with increasing $i$ because the line
is predominantly broadened by rotation 
for all inclinations, except for the lowest inclination where
the projected rotational velocity is lower than the expansion
velocity. The inclination angle affects also the position
of the red edge of the absorption which is at the line center
for $i=5^o$ and moves to the red as $i$ increases for $i<85$.
The fact that , for $i=55^o$ and $70^o$, 
the lines have two maximum absorptions
almost equally shifted from the line center to the blue and red
shows also that the rotation dominates over expansion in shaping the line
profile.
Additionally, for $i=85^o$, the line is in emission and is double-peaked
as expected for a rotating flow (e.g., Rybicki \& Hummer 1983).
The scattered line emission is very weak for all inclinations
except the highest inclination for which the background continuum
due to the disk photosphere is significantly reduced by foreshortening.
Generally, the line profiles of model~A show that indeed 
the wind is weak in the sense that the wind expansion velocity
is low compared to the rotational velocity 
and the mass loss rate is low.

\subsection{A strong steady wind case}

An example of a strong steady state wind is Model~C where  
$\MDOT_a=\pi\times10^{-8}~\MSUNYR$ and the white dwarf radiates at the same
rate as the disk $x=1$. The gross properties of the wind are
$\MDOT_a=2.1\times10^{-11}~\MSUNYR$, $\omega=32^o$ and
$v_r(10~r_\ast)= 3500~{\rm km~s^{-1}}$, the latter 
being comparable to $v_o$.  Note that a factor of $2\pi$ increase in the
system luminosity is accompanied by a factor of almost 400 increase in
mass loss (see table~1).

The line profiles for a strong steady state wind (fig.1k-1o, i.e. panels 
in the second from the bottom row) are markedly different from those for 
the complex weak model. The most striking difference is for high inclination
angles. For model~A, the line  develops a strong, more or less symmetric, 
double-peaked emission when seen nearly edge-on whereas for model~C, the
line  develops a typical P~Cygni profile with a blueshifted absorption and 
a redshifted emission. Additionally, for the strong steady wind, the line
is significantly broader than the weak complex wind at all inclinations. 
The rotation of the wind does not dominate over expansion in shaping the line.
In particular, the shape and strength of the blueshifted absorption is due
to expansion. For $i\simless 55^o$ there is a redshifted absorption that is 
broadened by the rotation. The redshifted absorption is weaker and narrower 
than the expansion-dominated blueshifted absorption.  We note a deep narrow 
absorption at $v\sim - 1000~{\rm km~s^{-1}}$ for $i=70^o$ (Fig. 1n)
and at $v\sim - 200~{\rm km~s^{-1}}$ for $i=85^0$ (Fig. 1o).
For $i<70^o$, line profiles for  model~A and C are similar in the respect that
the scattered emission is weak in both. At $i=85^o$,
the scattered emission is seen as weaker by the observer for the strong 
steady model than for the weak complex model because the background continuum
emission is stronger in the steady model. 
At this high inclination, the background continuum of Model~C
is dominated by the white dwarf contribution, $f_\ast \equiv 1- f_D=0.88> f_D$
(see Table~1). Thus, in contrast with Model~A or any model with x=0,
the Model~C background continuum is not much reduced by 
the foreshortening effect.

\section{Discussion and comparison with observations}

Some of the results shown in the previous section follow from the 
character of the hydrodynamic models (PSD~98, PSD~99) in a straightforward way.
For example, the lack of a strong ordered outflow in models 
without a white dwarf continuum (model A in the previous section) 
is consistent with line profiles showing weak, narrow absorption at low 
inclination and rotation dominated emission at high inclination.
Conversely, the models with a white dwarf at least as luminous as the disk
produce an ordered radial flow and therefore show broad blueshifted
absorption for a range of viewing angles. 

A key parameter shaping the total line profile -- made up of both 
scattered emission and absorption -- is the ratio of the expansion velocity to
the rotational velocity.  This can account for the differences in the line 
profile between models A and B and models C and D. For models with the white 
dwarf radiation switched on (C and D), the winds are more equatorial and so 
the rotational velocity decreases along the wind streamlines faster than for 
models with the white dwarf radiation switched off (A and B). This simple 
change in the wind geometry reduces the rotational velocity of the flow 
compared to the expansion velocity. The relatively higher expansion velocity 
has an important consequence on the scattered emission: for  models~C and D 
at high inclination  ($i\simgreat 60^o$), the red component of the
scattered emission becomes stronger -- stronger than the blue component of 
the emission and stronger than the blueshifted absorption so it is strong 
enough for  the total line to have a  P~Cygni profile (Fig. 1n, 1s and 1t).  
This also helps us to understand the complex shape of model C at $i$=70$^o$ 
(Figure 1n) as being a transition between the more familiar P-Cygni shape 
at that inclination of model D (Figure 1s) and the shape affected by scattered 
blueshifted emission of model B (Figure 1i).

Perhaps the most striking prediction of our calculations is the double-humped
structure near line center that persists even in model C and D at intermediate
inclinations. We identify this structure with a non-negligible redshifted 
absorption which is formed in the slow wind where the rotational velocity 
dominates over expansion velocity.  Figure~2 compares the line profiles 
calculated for model~C taken from Figure~1 (Fig. 1k-1o) with line profiles 
also calculated for model~C but with the assumption that the slow wind 
contributes nothing to the profile near the equator.  This change is achieved
by setting the ion fraction, 
$\xi_{ion}=0$ for those $\bar {\omega}$ and $z$, 
that satisfy the condition  $\bar {\omega} > 4 r_\ast$ and
$z < (\bar {\omega}-4 r_\ast)~tan~30^o$, where
$\bar {\omega}$ and $z$ are the radius and height
cylindrical coordinates (see Fig. 3c in
PSD~99 - note that our $\bar {\omega}$ corresponds to $r$ on that figure).  
As one may expect, the model
with the transparent slow wind (solid lines) predicts much weaker redshifted
absorption and, as a result, the double-humped structure near line center 
has almost completely disappeared.
In particular, the line profile for $i=70^o$ (Fig.~2d)
looks rather more like the smooth P~Cygni profile typical of pure
expansion.

When compared with the properties of observed line profiles, our 
models show some features which appear to be qualitatively consistent. 
These include the appearance of the classical P-Cygni shape for a range of 
inclinations, the location of the maximum depth of the absorption 
component at velocities less than the terminal velocity (a requirement 
discussed in Drew 1987), and the transition from net absorption to net 
emission with increasing inclination.  Whilst this much is encouraging,
it is important to move onwards to a more critical, and direct, comparison 
with spectrally well-resolved observational data.  To this end we can
make use of UV spectroscopy described in full by Hartley et al. 
(in press).  In Figure 3, we show a comparison between profiles derived
from model C with a transparent slow wind and 
observations of the C{\sc iv}~1549\AA\ and Si{\sc iv} 1403\AA\ transitions in 
the spectrum of the brightest nova-like variable, IX~Vel. We show profiles 
synthesized for $i = 60^o$ as this is the inclination believed to 
be appropriate to this binary (Beuermann \& Thomas 1990).  

In Figure 3, there is clearly too little absorption or emission equivalent 
width in the synthetic profiles, when compared with the C{\sc iv} profile.  
This certainly implies too low a model mass loss rate.  This is not a 
surprise in that the impression emerging from previous work is that mass loss 
rates in excess of $1 \times 10^{-10}$ M$_{\odot}$ yr$^{-1}$ are likely to be 
required (e.g., Drew 1987; Mauche \& Raymond 1987; Shlosman \& Vitello 1993;
Knigge, Woods \& Drew 1995; Mauche \& Raymond 2000): 
the mass loss rate for the model shown is just $2.1 \times 10^{-11}$ 
M$_{\odot}$ yr$^{-1}$.  It is also no surprise that the observed C{\sc iv} 
profile exhibits more redshifted emission than the $i = 60^o$ model 
prediction -- this is also a problem that has been apparent for a while 
(see e.g. Mauche, Lee \& Kallman 1997; Ko et al. 1998 and below).

A more intriguing difficulty is encountered on comparing the observations
with the profile synthesized for the flow comprising both the fast stream and 
the slow wind (dashed line): there is a residue of a rotation-induced double 
minimum in the latter not present in the former.  The structure that is 
apparent near line center in the observed C{\sc iv} line profile is not due 
to rotation -- it arises from the doublet nature of the transition (the fine 
structure splitting is 484 km s$^{-1}$, referred to the blue component).  
The doublet responsible for the Si{\sc iv} 1397\AA\ feature is so 
well-separated ($\Delta v \sim 1930$ km~s$^{-1}$) that the red component at 
1403\AA\ in IX~Vel's spectrum is a de facto singlet. In Figure 3, the 
blueshifted absorption in the Si{\sc iv} profile reassuringly resembles that 
in C{\sc iv}, reaching a minimum at the same projected velocity and seemingly 
sampling the same basic kinematics.  Whilst there is a narrow, very-slightly 
redshifted absorption at $+$100 km~s$^{-1}$ in the Si{\sc iv} profile, it 
lacks a counterpart in the blue component of the line and appears to be 
rather too sharp to attribute to rotation as in the model profiles (see 
Figure~1 and 2).  In making this comparison we have not needed to be careful 
in our selection of observations -- this absence of a rotational effect from 
observed line profiles appears to be the norm (e.g., Cordova \& Mason 1984; 
Drew 1990; Mason et al. 1995; Prinja \& Rosen 1995; Mauche, Lee \& 
Kallman 1997; Prinja et al. 2000a; b).

By making the slow wind transparent, as in the synthesis shown as the solid
line in Figure 3, the double minimum apparent near line center is replaced
by a more plausible single, blueshifted minimum.  This is rather more
satisfactory, when compared with the observations.  

The significance of this discrepancy and its cure seems clear in one respect: 
the wind-formed line profiles seen at ultraviolet wavelengths cannot originate
in a flow where rotation and poloidal expansion are comparable.  The UV lines 
must trace gas that expands substantially faster than it rotates.  This 
criterion is more likely to be satisfied by angular momentum conserving 
radiation driven disk winds like those considered here than by 
constant angular velocity flows (as might be predicted for MHD-driven
winds).  Accordingly, wholesale abandonment of radiation-driven disk wind
models in favour of pure (co-rotating) MHD disk winds could actually make 
matters worse, rather than better.  Regardless of the details of the driving 
mechanism, the comparison we have made here between model and observation 
indicates that the seat of the UV wind-formed line profiles cannot be a
wind that is both slowly expanding and rapidly rotating.  

It is important to note that many of the details of our results depend 
on details of the assumptions about the disk and microphysics of the wind.
For example, the assumed form for the background radiation field inevitably 
plays a role in the comparison of observed profiles with synthetic profiles 
based on any model, kinematic or dynamical.  However for line fitting based 
on radiation-driven wind models, the adopted disk and white dwarf radiation 
fields are even more important because they determine all wind properties
except the initial Keplerian component of motion.  In our calculations, we 
followed PSD~99 and adopted the dependence of the disk radiation on radius 
according to the standard steady state disk model (e.g., Pringle 1981).
This assumption is a good starting point but it should be remembered that 
actual disks may only be very crudely  described by simple theory 
or they may yield unexpected features such as chromospheric emission.  In 
fact, spectral synthesis models of accretion disk photospheres (e.g., 
Linnell \& Hubeny 1996; Wade \& Hubeny 1998; Wade \& Orosz 1999) have 
typically failed to adequately reproduce observed energy distributions where 
direct comparisons have been made (see e.g. Long et al. 1994).  There has 
been a similar lack of success in past calculations of accretion disk line 
emission.  Specifically, the accounting for both the strengths of and the 
flux ratios among various emission lines in a large ensemble of observed CVs 
has not been compelling (Mauche, Lee \& Kallman 1997; Ko et al., 1998).
These models either miss a crucial physical component or employ an 
inappropriate physical assumption which affects the predicted line emission.  
It is for this reason that, in comparing model wind profiles 
with observation, we concern ourselves more with the absorption
component than with the emission.

In our line-profile calculations we also assumed that the ionization fraction 
of scattering species is constant ($\xi_{ion}$ is set to unity). A proper
allowance for a variation of $\xi_{ion}$ with position to values less
than one can only serve to weaken the overall line profile.  (Examples of 
plots of this potentially very marked positional dependence may be found in 
the work of Shlosman \& Vitello 1993).  

Line-driven disk wind models predict that winds driven from an accretion 
disk without a strong central star yield complex, unsteady outflow (PSD~98 
and PSD~99). Therefore we might expect $\simless {\rm 100~km~s^{-1} }$ fine 
structure in blueshifted absorption features and that this fine structure is 
time-variable. Our results show that line profiles for such complex outflows 
(model~A and model~B) are relatively smooth, and that whatever structure is 
present in the profile at velocities $\leq 100$ km s$^{-1}$ is small compared 
with the net absorption on larger velocity scales.  This can be attributed 
partially to the fact the line absorption is nearly saturated in model B (and 
the non-zero residual flux is due to filling in by emission), and also to the 
more general point that the density and velocity fluctuations in these wind 
models are just too fine in spatial scale to imprint on the line profile.

To bridge the gap between the kinematics of the radiation driven models 
presented here and reality, we will probably need to understand why the wind 
seen at UV wavelengths should expand with velocities that at least match
the rotational velocity over a wide range of radii.  There are a few 
possible changes to the kinematics of the radiation driven models that can 
improve the situation.  For example, it might be that the slow wind present 
in PSD~99's models would not contribute to the UV spectrum
because its ionization stage is inappropriate. This possibility
still requires an increase in the mass loss rate in the fast stream. 
Another possibility, which we think is more likely, is that the wind
launched at large radii is not slow as PSD~99 predicted but some 
mechanism quickly accelerates the expansion to a velocity exceeding
the local rotation speed.  It already seems necessary to bring in e.g.,  
MHD processes to yield higher rates of mass loss (see e.g., Mauche \& Raymond 
2000; Drew \& Proga 2000; Proga 2000).  Now we can place a more specific 
requirement on the as-yet unidentified additional driving: it should 
accelerate dense material launched by the radiation force at larger radii to 
an expansion velocity at least comparable to the rotational velocity on a 
length scale of several white dwarf radii. Increasing the mass loss rate
by increasing the wind density, rather than the poloidal velocity,  
is unworkable because such a wind would produce rotational features even
stronger than those obtained here.

ACKNOWLEDGMENTS: 
The work presented in this paper was performed while DP held
a National Research Council Research Associateship at NASA/GSFC.
Computations were supported by NASA grant NRA-97-12-055-154.  This
paper makes use of observations by the NASA/ESA Hubble Space
Telescope, obtained at the Space Telescope Science Institute, which
is operated by the Association of Universities for Research in
Astronomy, Inc., under NASA contract NAS 5-26555.

\newpage
\section*{ REFERENCES}
 \everypar=
   {\hangafter=1 \hangindent=.5in}

{

  Batchelor G.K. 1967, An Introduction to Fluid Mechanics (Cambridge:
  Cambridge University Press)

  Boggess et al. 1978, Nature, 275, 37

  Beuermann, K., Thomas, H.-C., 1990, A\&A, 230, 326

  Castor, J.I., Abbott, D.C.,  Klein, R.I. 1975, ApJ, 195, 157 (CAK)
  
  Cannizzo J.K. 1993, in {\it Accretion Disks in Compact Stellar Systems}, ed.
                   J.C. Wheeler, Singapore, World Scientific Publishing, p.6

  Cordova, F.A.,  Mason K.O. 1982, ApJ, 260, 716

  Cordova, F.A.,  Mason K.O. 1984, MNRAS, 206, 874

  Cordova, F.A.,  Mason K.O. 1985, ApJ, 290, 671

  Drew, J.E. 1987, MNRAS, 224, 595

  Drew, J.E. 1990, in {\it Physics of Classical Novae}, Lecture Notes
       in Physics 369, Springer-Verlag, Berlin, p. 228

  Drew, J.E. \& Verbunt, F. 1985, MNRAS, 213, 191

  Feldmeier, A., Shlosman, I. 1999, ApJ, 526, 344 

  Feldmeier, A., Shlosman, I.,  Vitello, P.A.J 1999, 526, 357

  Heap, S. R.  et al. 1978, Nature, 275, 385

  King A.R., Frank, J., Jameson, R.F., Sherrington, M.R. 1983, MNRAS, 203, 677

  Knigge, C.,  Drew, J.E. 1997, ApJ, 486, 445

  Knigge, C., Woods, J.A., Drew, J.E. 1995, MNRAS, 273, 225

  Ko, Y., Lee, P. Y., Schlegel, E. M., \& Kallman, T. R. 1996, ApJ, 457, 363 

  Krautter J., Vogt N., Klare G., Wolf B., Duerbeck H. W., Rahe J., 
  Wargau W., 1981, A\&A, 102, 337

  Linnell, A.P., \& Hubeny, I. 1996, ApJ, 471, 958

  Long, K.S., Wade, R.A., Blair, W.P., Davidsen, A.F., \& Hubeny, I.
     1994, ApJ, 426, 704

  Marti, F. \& Noerdlinger P.D. 1977, ApJ, 215, 247

  Mason, K. O., Drew, J. E., Cordova, F. A., Horne, K., Hilditch, R., 
  Knigge, C., Lanz, T., \& Meylan, T. 1995, MNRAS, 274, 271

  Mauche, C. 1998, in ASP Conf. Ser. 137, Wild Stars in the Old West:
  Proceedings of the 13th North American Workshop on Cataclysmic Variables
  and Related Objects, ed. S. Howell, E. Kuulkers, \& C. Woodward (San
  Francisco: ASP), 113

  Mauche, C.W., Lee, Y.P., Kallman, T.R. 1997, ApJ, 477, 832

  Mauche C.W., Raymond J.C. 1987, ApJ, 323, 690

  Mauche C.W., \& Raymond J.C. 2000, ApJ, 323, 690

  Mauche, C. W., Raymond, J. C., Buckley, D. A. H., Mouchet, M., 
  Bonnell, J., Sullivan, D. J., Bonnet-Bidaud, J.-M., \& Bunk, W.H. 
  1994, ApJ, 424, 347

  Naylor, T., Bath, G. T., Charles, P. A., Hassall, B. J. M., Sonneborn, 
  G., van der Woerd, H., \& van Paradijs, J. 1988, MNRAS, 231, 237

  Owocki, S.P., Castor J.I., \& Rybicki, G.B. 1988, ApJ, 335, 914

  Pereyra, N.A., Kallman, T.R., \& Blondin, J.M. 1997, ApJ, 477, 368

  Pereyra N.A., Kallman T.R.,  Blondin J.M., 2000, ApJ, 532, 536

  Pringle J.E. 1981, ARAA, 19, 137

  Prinja, R.K., Knigge, C., Ringwald, F.A., Wade, R.A. 2000, MNRAS, 318,
  368, b

  Prinja, R.K., Ringwald, F.A., Wade, R.A., Knigge, C. 2000, MNRAS, 312,
  316, a

  Prinja, R.K., Rosen S.R. 1995, MNRAS, 273, 461

  Proga, D. 1999, MNRAS, 304, 938

  Proga, D 2000, ApJ, 538, 684

  Proga, D., Stone J.M., \& Drew J.E. 1998, MNRAS, 295, 595 (PSD~98)

  Proga, D., Stone J.M., \& Drew J.E. 1999, MNRAS, 310, 476 (PSD~99)

  Rybicki G.B., \& Hummer D.G.  1978, ApJ, 219, 654

  Rybicki G.B., \& Hummer D.G.  1983, ApJ, 274, 380

  Shakura N.I., \& Sunyaev R.A. 1973 A\&A, 24, 337

  Shlosman I., \& Vitello P.A.J. 1993, ApJ, 409, 372

  Shlosman I., Vitello P.A.J., \& Mauche C. W. 1996, ApJ, 461, 377

  Sobolev, V.V. 1957, Soviet Astron.-AJ, 1, 678

  Wade, R.A., \& Hubeny, I. 1998, ApJ, 509, 350

  Wade, R.A., \& Orosz J.A. 1999, ApJ, 525, 915

  Woods J.A., Verbunt, F., Collier Cameron, A., Drew, J.E., \&
  Piters, A. 1992, MNRAS, 255, 237

}

\eject

\newpage
\centerline{\bf Figure Captions}

\vskip 4ex
\noindent
Figure~1 -- Line profiles for hydrodynamical disk wind models as a function
of inclination angle, $i$. The values of $i$ are $5^o$, $30^o$, $55^o$, $70^o$ 
and $85^o$ (panels in first, second, third, fourth and fifth column,
respectively). 
The panels in the top and second from the top row
, a-e and f-j, are both for  models with $x=0$ but with $\MDOT_{a} = 
10^{-8}~\MSUNYR$ (model A) and 
$\MDOT_{a} = \pi\times 10^{-8}~\MSUNYR$ (model B), respectively.  
The panels in the second from the bottom and the bottom row, k-o and p-t, 
show results for models both with $\MDOT_{a} = \pi \times 10^{-8}~\MSUNYR$, but
with $x=1$ (model C) and $x=3$ (model D).  The panels for model~A
and model~B (a-e and f-j) show the
effect on the profiles of increasing the disk luminosity alone,
while the panels for model~B, model~C and D (f-j, k-o, and p-t)
show the effect of adding in an 
increasingly larger stellar component ($x~=~$ 0, 1 and 3) to the radiation
field.  The total systems luminosity and therefore the wind mass
mass loss rate increase from  top to  bottom.
The choice of input  parameters for the hydrodynamical models
shown is the same as in Figure~10 of PSD~98 and in Figure~2 in PSD~99.
The zero velocity corresponding to the line center is indicated by the vertical
line.
Note the difference in the velocity and flux ranges in the planes
for $i=85^o$ (fifth column).

\vskip 4ex
\noindent
Figure~2 -- Line profiles for hydrodynamical disk wind model~C as a function 
of inclination angle, $i$. The values of $i$ are as in Figure~1. The figure 
compares the line profiles for model~C from Figure~1 (Fig.~1k-1o), 
dashed lines here,
with the line profiles for model~C with the slow wind material between 
the disk equator and the 'fast stream' assumed to be optically thin 
in the line, solid lines. We define the location of  the transparent wedge 
in the wind using the cylindrical coordinates, $\bar {\omega}$ and $z$, 
that satisfy the condition  $\bar {\omega} > 4 r_\ast$ and
$z < (\bar {\omega}-4 r_\ast)~tan~30^o$ (see the text for more details).

\vskip 4ex
\noindent
Figure~3 -- 
Comparison between profiles derived
from model C with and without the assumption that a slow wind 
is transparent (thick solid and thick dashed line, respectively) and 
observations of the C{\sc iv}~1549\AA\ and Si{\sc iv} 1403\AA\ transitions in 
the spectrum of the brightest nova-like variable, IX~Vel (Hartley et al.,
in press).  The synthetic profiles 
are  for the inclination of $60^o$. For clarity, we  renormalized 
the synthetic profiles so their continuum flux is at 1.25 not at 1.00 as
the continuum flux of the observed profiles.

\newpage

\begin{figure}
\begin{picture}(180,590)

\put(280,423){\includegraphics{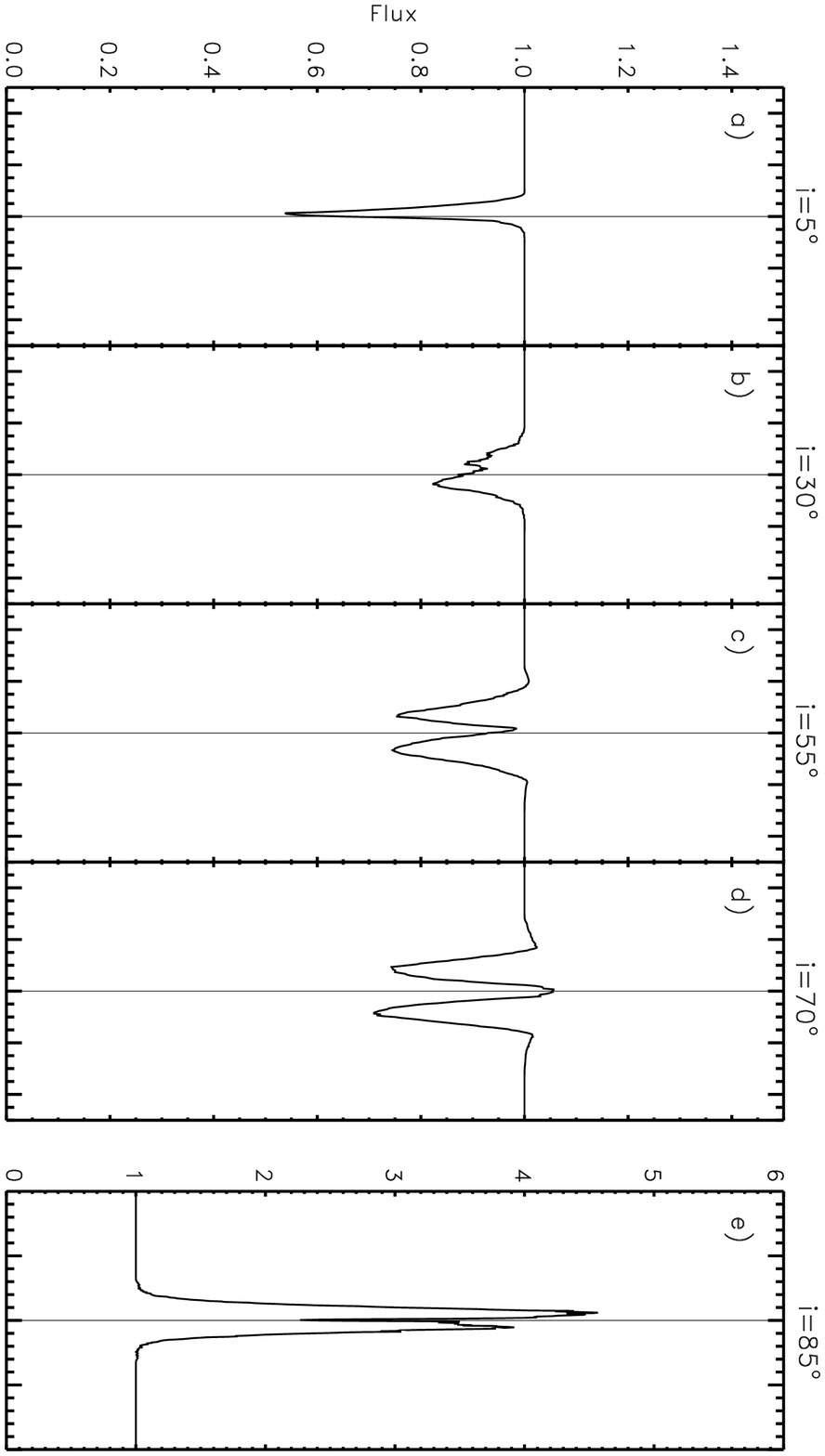}}

\put(280,282){\includegraphics{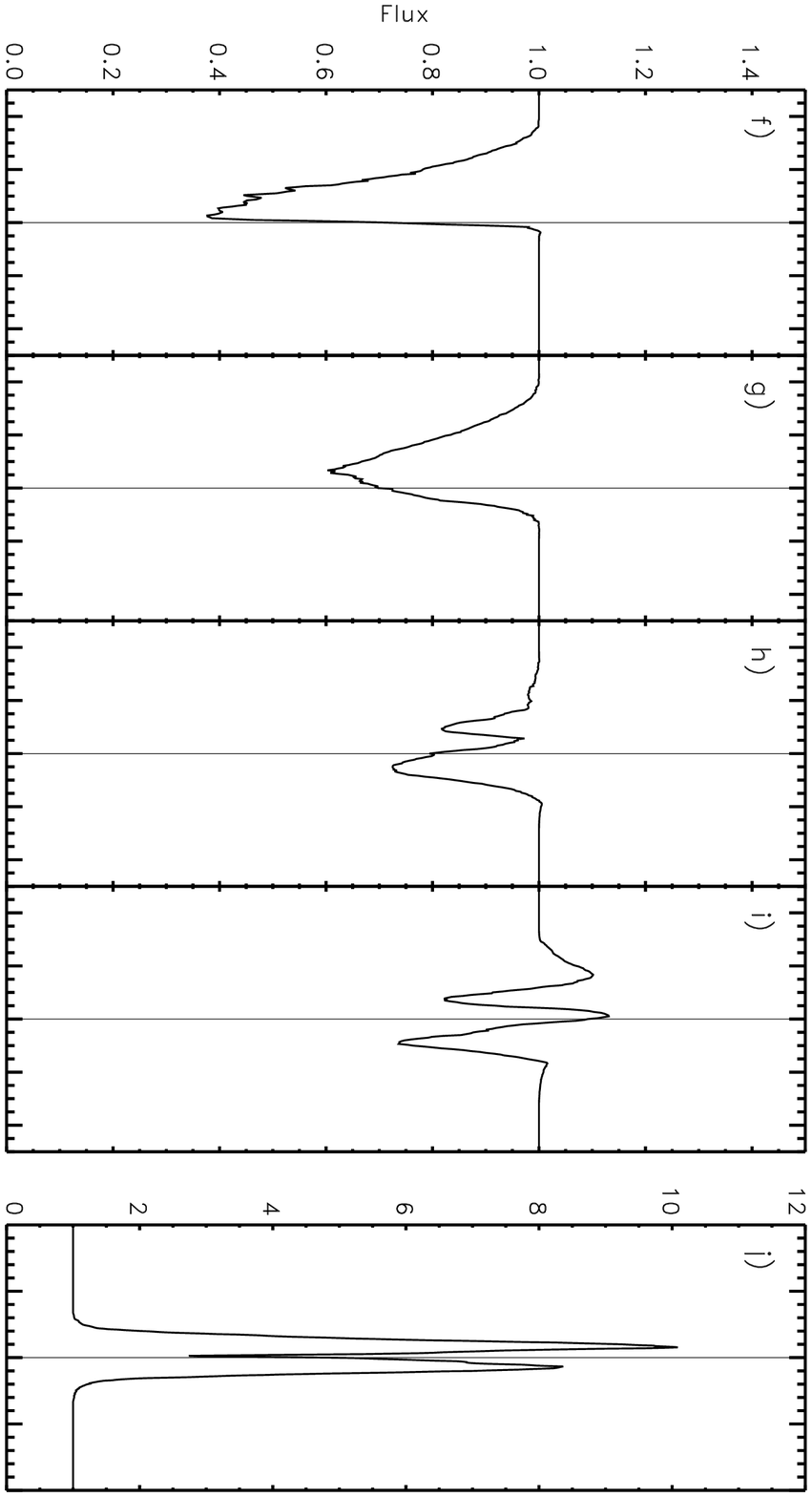}}

\put(280,141){\includegraphics{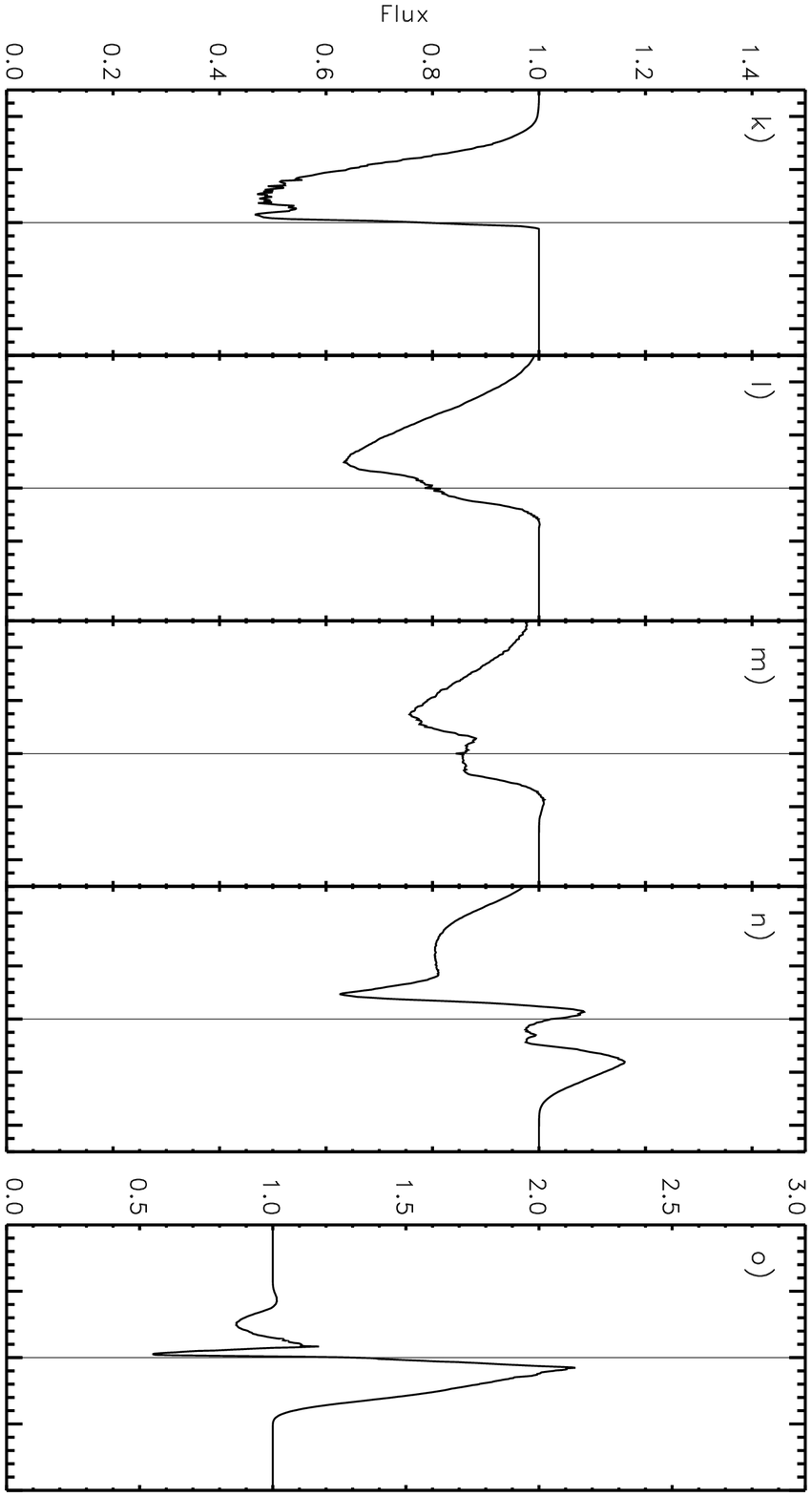}}

\put(280,0){\includegraphics{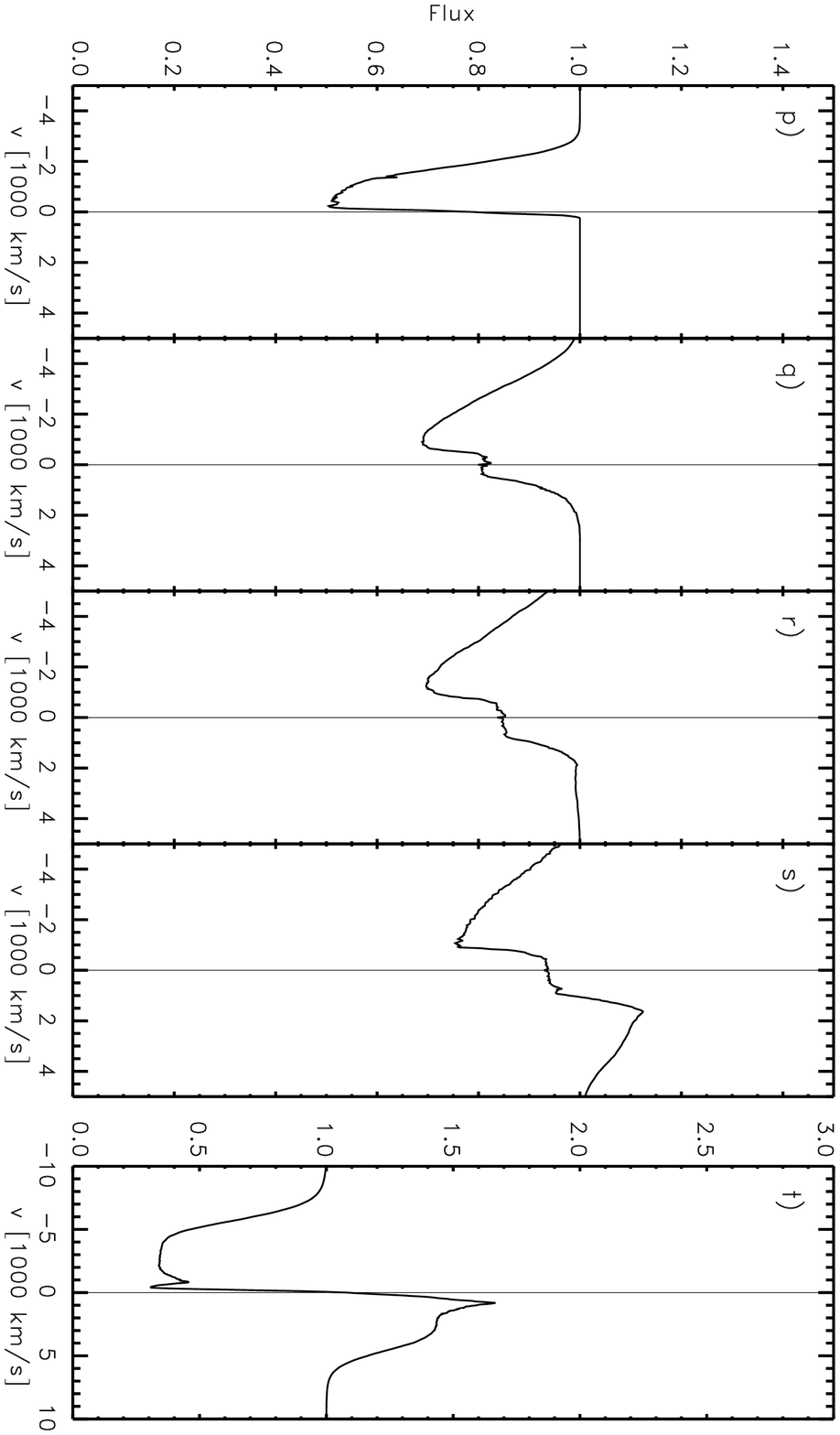}}

\end{picture}
\caption{}
\end{figure}

\begin{figure}
\begin{picture}(180,590)

\put(280,423){\includegraphics{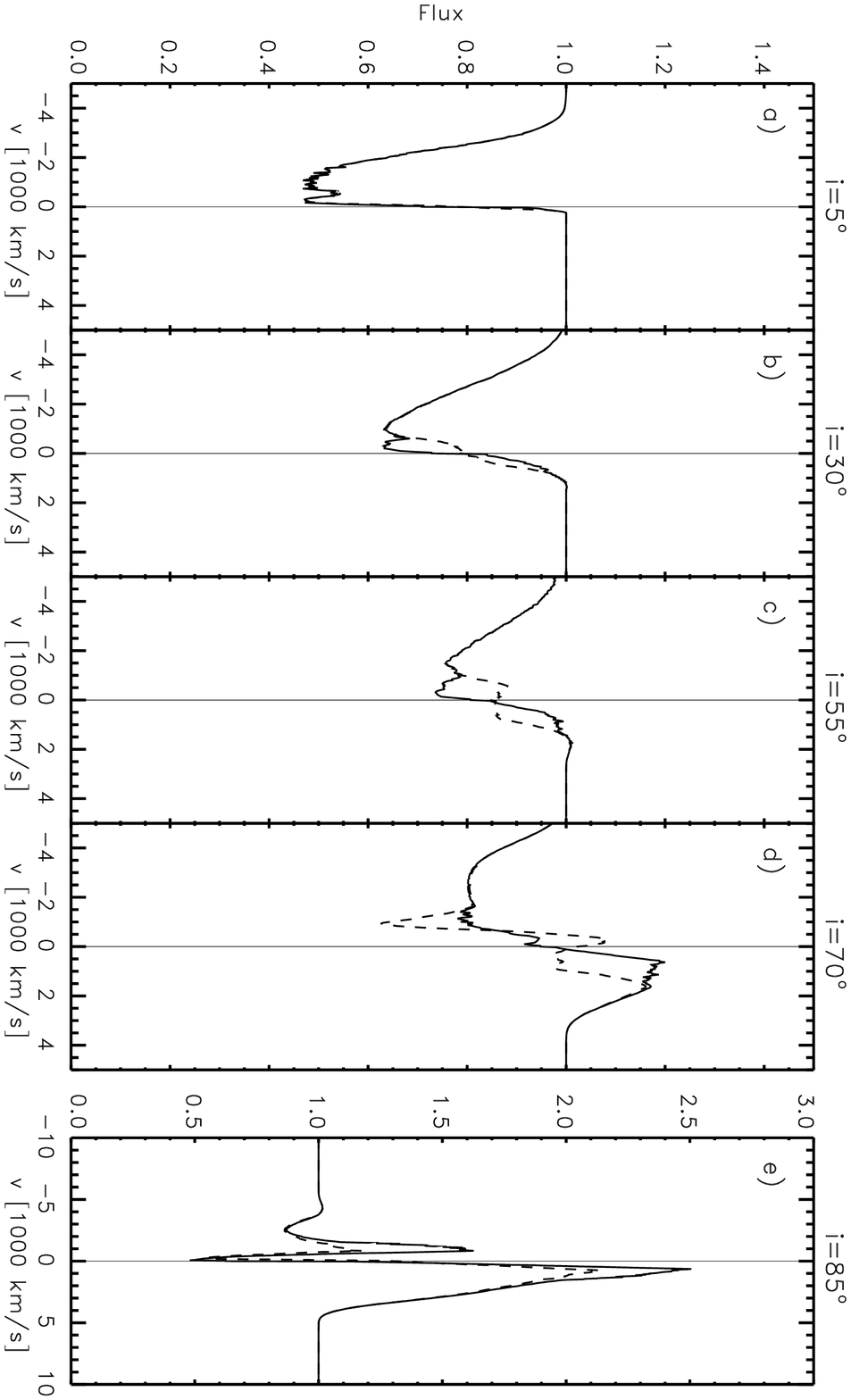}}

\end{picture}
\caption{}
\end{figure}

\begin{figure}
\begin{picture}(180,590)

\put(280,423){\includegraphics{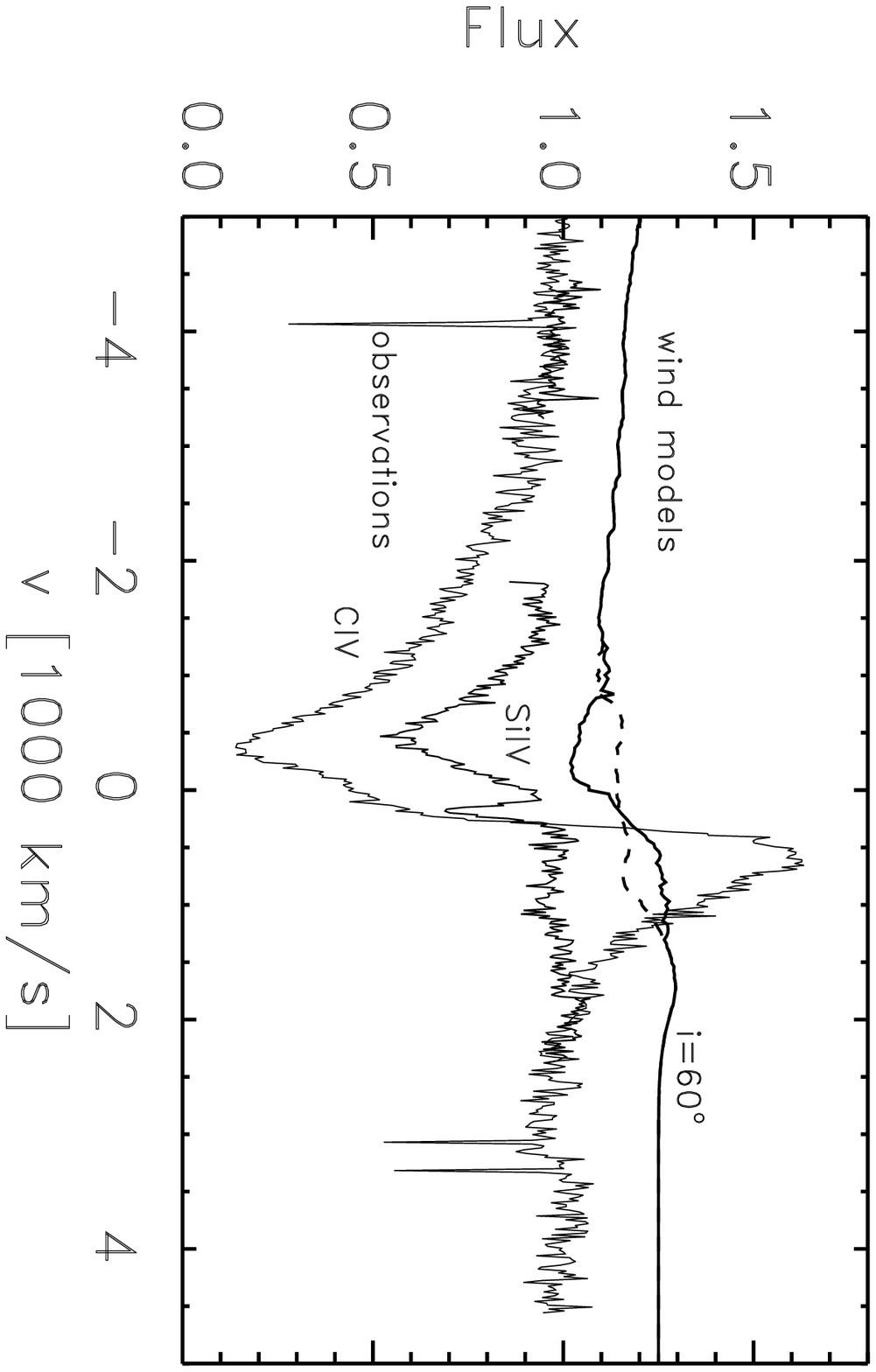}}

\end{picture}
\caption{}
\end{figure}

\eject

\begin{table*}
\footnotesize
\begin{center}
\caption{ Input parameters and main characteristics of the hydrodynamical models used in this paper
(see also PSD~99)}
\begin{tabular}{c c c r c c c c c r c   } \\ \hline
     &                         &   &                    & & &     & & & &                              \\
 model & $\MDOT_a$          & x & $L_{tot}$ &   $f_D(i=30^o)$
  & $f_D(i=55^o)$ & $f_D(i=70^o)$ 
& $f_D(i=85^o)$ &  
 $\MDOT_w$  &   $v_r(10 r_\ast)\ $ & $\omega$   \\
     & (M$_{\odot}$ yr$^{-1}$) &   & $\LSUN$  & & & & & (M$_{\odot}$ yr$^{-1}$) 
  & $(\rm km~s^{-1})\  $ & degrees    \\ \hline

     &                         &   &                         & &  & & & &                    \\
A    &$  10^{-8}$              & 0 & 7.5   & 1.00 & 1.00  & 1.00 &  1.00 & $ 5.5\times10^{-14}$    & 900           \
      &  50  \\
B    &$ \pi \times 10^{-8}$    & 0 & 23.4  & 1.00 & 1.00  & 1.00 & 1.00 & $ 4.0\times10^{-12}$    & 3500         \
       &  60  \\
C    &$ \pi \times 10^{-8}$    & 1 & 46.9  & 0.78 & 0.77 & 0.71 & 0.12 &  $ 2.1\times10^{-11}$    & 3500          \
      &  32  \\
D    &$ \pi \times 10^{-8}$    & 3 & 93.8  & 0.63 & 0.61 & 0.56 & 0.08 & $ 7.1\times10^{-11}$    & 5000          \
      &  16  \\
\hline

\end{tabular}
\end{center}
\end{table*}
\eject

\end{document}